\documentclass[twocolumn,showpacs,preprintnumbers,amsmath,amssymb]{revtex4}
\usepackage{amsmath}
\usepackage{amsfonts}
\usepackage{amsthm}
\usepackage{amssymb}
\usepackage{graphicx}
\usepackage{mathrsfs}
\usepackage{latexsym}
\usepackage{graphics}

\begin{document}

\title{Cavitation Bubble Dynamics inside Liquid Drops in Microgravity}

\author{D. Obreschkow$^{1,2}$}
\author{P. Kobel$^{1,3}$}
\author{N. Dorsaz$^{1,4}$}
\author{A. de Bosset$^1$}
\author{C. Nicollier$^5$}
\author{M. Farhat$^1$}

\affiliation{
$^1\,$ Laboratoire des Machines Hydrauliques, EPFL, 1007 Lausanne,
Switzerland\\
$^2\,$Physics Department, Oxford University, Oxford, OX1 3PU, UK\\
$^3\,$Max Planck Institute for Solar System Research, 37191 Katlenburg-Lindau,
Germany\\
$^4\,$Institut Romand de Recherche Num\'{e}rique en Physique des
Mat\'{e}riaux, EPFL, 1015 Lausanne, Switzerland\\
$^5\,$ESA European Astronaut Centre, Cologne, Germany, and NASA Johnson
Space Center, Houston, TX}


\date{\today}

\begin{abstract}
[Almost identical to PRL 97, 094502 (2006)] We studied spark-generated
cavitation bubbles inside water drops produced in microgravity. High-speed
visualizations disclosed unique effects of the \emph{spherical} and nearly
\emph{isolated} liquid volume. In particular, (1) toroidally collapsing bubbles
generate two liquid jets escaping from the drop, and the "splash jet" discloses
a remarkable broadening. (2) Shockwaves induce a strong form of secondary
cavitation due to the particular shockwave confinement. This feature offers a
novel way to estimate integral shockwave energies in isolated volumes. (3)
Bubble lifetimes in drops are shorter than in extended volumes in remarkable
agreement with herein derived corrective terms for the Rayleigh-Plesset
equation.
\end{abstract}

\maketitle

\emph{Introduction} $-$ Hydrodynamic cavitation is a major source of erosion
damage in many industrial systems, such as fast ship propellers, cryogenic
pumps, pipelines, and turbines \cite{Brennen 95}. Yet, \emph{controlled}
cavitation-erosion proves a powerful tool for modern technologies like
ultrasonic cleaning \cite{Song 04}, effective salmonella destruction
\cite{Wrigley 92}, and treatment for kidney stones \cite{Pishchalnikov 03}.
Such erosion is associated with liquid jets and shockwaves emitted by
collapsing cavitation bubbles, but the relative importance of these two
processes remains a topic of debate \cite{Shima 97}. Fundamental understanding
requires studies of single bubbles in different liquid geometries, since bubble
dynamics strongly depends on nearby surfaces by means of boundary conditions
imposed on the surrounding pressure field \cite{Plesset 71,Brennen 95}. Recent
investigations revealed interesting characteristics of bubbles collapsing next
to flat \cite{Brujan 02} and curved \cite{Tomita 02} rigid surfaces or flat
free surfaces \cite{Robinson 01, Pearson 04}. It is thus a promising idea to
study bubbles inside \emph{spherical drops} and probe their interaction with
closed spherical free surfaces. However, for centimeter-sized volumes, such
geometries are inaccessible in the presence of gravity and require a
microgravity environment, even though gravity plays a negligible \emph{direct}
role for most single bubble phenomena.

In this letter, we present the first experimental study of single bubble
dynamics inside centimetric quasi-spherical water drops produced in
microgravity. High-speed imaging revealed key-implications of nearly
\emph{isolated}, \emph{finite} and \emph{spherical} liquid volumes on bubble
evolution and subsequent phenomena. The results will be presented in three
steps covering (1) jet dynamics, (2) shockwave effects and (3) spherical
collapse.

\emph{Experimental Setup} (for a detailed description and figures see
\cite{FS}) $-$ Microgravity was achieved in manned parabolic flights conducted
by the European Space Agency ESA (42nd parabolic flight campaign). The nominal
flight manoeuvres offer sequences (i.e.~parabolas) of 20 s of weightlessness
with a residual g-jitter of 0.02-0.05 g at typical frequencies of 1-10 Hz. Our
experiment and all calibrations were carried out on 93 parabolas on three
consecutive flight days.

\begin{figure}[h]
  \includegraphics[width=8.3cm]{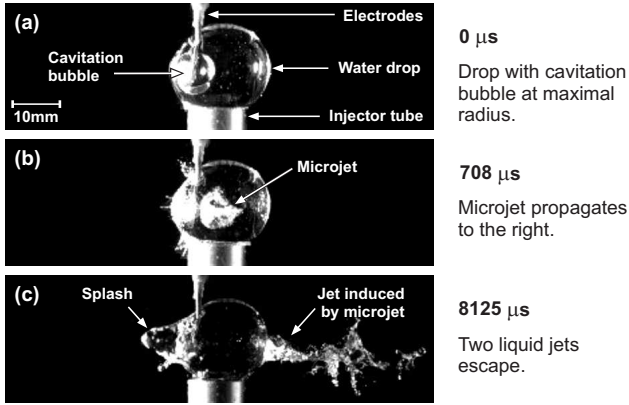}
  \caption{Two liquid jets generated by the toroidal implosion of
  the a cavitation bubble ($\alpha$=0.3,$\varepsilon$=0.45).
  Movies at \cite{FS}.}
  \label{fig 1}
\end{figure}

Each parabola was used to run \emph{one} fully automated experimental cycle,
consisting in the computer-controlled growth of a centimetric water drop and
the generation of a single bubble inside the drop. The liquid was expelled
though an \emph{injector tube}, specially designed to fix and stabilize the
ensuing truncated drop (Fig.\ \ref{fig 1}). The bubble was generated through a
spark-discharge between two thin electrodes immersed in the drop. A high-speed
CCD-camera (Photron Ultima APX, up to 120\,000 frames/s) recorded the fast
bubble dynamics for a sequence of 11 ms. Three parameters were independently
adjustable: (1) the initial drop size [radius $R_{d,min}$ = 8-13 mm] by
changing the expelled water volume, (2) the maximal bubble size [radius
$R_{b,max}$ = 2-10 mm] by altering the discharge energy between 8 and 1000 mJ,
(3) the distance $d$ between drop center and bubble center by precisely
adjusting the electrodes position. We shall usually describe a particular
setting in the dimensionless parameter space ($\alpha,\varepsilon$), where
$\alpha\equiv R_{b,max}/R_{d,min}$ (\emph{relative bubble radius}),
$\varepsilon\equiv d/R_{d,min}$ (\emph{eccentricity}).

\emph{Jet Dynamics} $-$ As a first result, we overview the collapse of
eccentrically placed cavitation bubbles in the parameter range
($0.2<\alpha<0.6$, $0.3<\varepsilon<0.8$), recorded at 24\,000 frames/s
(extracts in Fig.\ \ref{fig 1}). The nearby free surface breaks the spherical
symmetry leading to a \emph{toroidal implosion} \cite{Robinson 01} with a fast
\emph{microjet} ($>$ 50 m/s) directed perpendicularly away from the closest
surface element (Fig.\ \ref{fig 1}b). This microjet accelerates the surrounding
volume forming a jet that escapes from the right surface in the figure at a
reduced velocity (6 m/s). In the meantime, a slower \emph{splash} escapes in
the opposite direction (Fig.\ \ref{fig 1}c). This double-jet picture is
consistent with established studies of bubbles in the vicinity of free surfaces
\cite{Crum 79,Robinson 01}, and provides the \emph{first direct visualization}
of both bubble-induced jets escaping from a steady liquid volume.

\begin{figure}[h]
  \includegraphics[width=8.3cm]{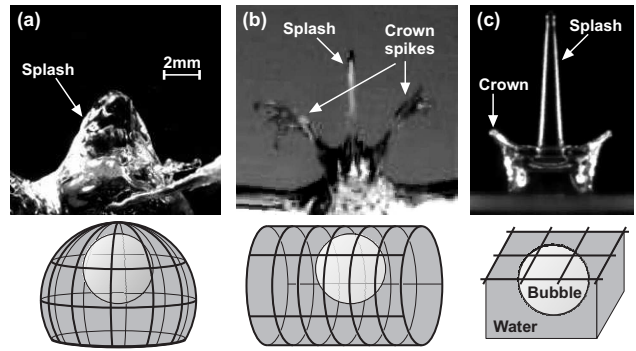}
  \caption{Splash on several free surfaces. (On (a), electrodes enter from
  the right. (b) was provided by E. Robert \cite{Robert 04}.)}
  \label{fig 2}
\end{figure}

Close investigation of the splash geometry revealed a remarkable and
reproducible diameter broadening compared to similar jets on ground-based
\emph{flat} free surfaces (Fig.~\ref{fig 2}a,c). In the latter case, the
narrowness is due to a very localized high-pressure peak between bubble and
free surface \cite{Robinson 01}. We believe that this pressure peak is much
broader beneath \emph{spherical} free surfaces, since the distance between
bubble boundary and surface varies more smoothly. Such a pressure zone
broadening suggests an explanation for a broader splash and will be subject to
forthcoming simulations. Additionally, we discovered that the typical
\emph{crown} surrounding the splash on flat surfaces (Fig.~\ref{fig 2}c) is
absent on the spherical surface (Fig.~\ref{fig 2}a). For comparison it is
enlightening to consider cylindrical free surfaces, which are straight
\emph{and} circular in two orthogonal directions. One might reason that such
surfaces exhibit a hybrid behavior between (a) and (c). Indeed, recent studies
of bubbles inside a falling water stream \cite{Robert 04} revealed a
non-circular crown consisting of two spikes in the straight direction of the
surface, while no crown seems present perpendicular to the figure
(Fig.~\ref{fig 2}b). These observations definitely disclose non-trivial
features in splash dynamics and crown formation and promise an interesting
avenue for future research.

\begin{figure}[h]
  \includegraphics[width=8.3cm]{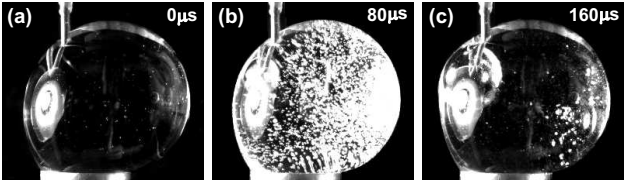}
  \caption{Example of Secondary cavitation: the primary shockwave radiated between (a) and (b) induces a
  short-lived \emph{mist of microbubbles}.
  Movies at \cite{FS}.}
  \label{fig 3}
\end{figure}

\emph{Shockwave Effects} $-$ Many high-speed visualizations contain isolated
frames showing a short-lived brilliant \emph{mist} (Fig.\ \ref{fig 3}). This
mist lasts for roughly 50 $\mu$s and is visibly composed of sub-millimetric
bubbles (here called \emph{microbubbles}). Strikingly, these mists exactly
coincide with the instants of predicted shockwave radiation. Namely, the
\emph{spark (or primary) shockwave} is emitted at the spark generation
initiating the bubble growth, and \emph{collapse (or secondary) shockwaves} are
radiated at the bubble collapse and subsequent collapses of rebound bubbles
\cite{Kodama 00}. Although these shockwaves were not directly observable with
the present setup, their exact synchronization with microbubbles clearly
discloses a causal relation. We claim that the mist of microbubbles is a strong
form of shockwave-induced \emph{secondary cavitation} \cite{Tomita 91}. In
other words, microbubbles are small cavitation bubbles inside the drop volume,
arising from the excitation of nuclei (i.e.~microscopic impurities and
dissolved gas) at the passage of shockwaves \cite{Wolfrum 03,Sankin 05}. That
the microbubbles were indeed cavities was confirmed by showing that their
size/lifetime ratio matches theoretical cavity life cycles.

The surprisingly high abundance of microbubbles compared to faint traces seen
in ground experiments \cite{Tomita 91} is reasonably explained by a unique
\emph{shockwave confinement} in the case of \emph{isolated liquid volumes}. The
shockwave can be reflected many times on the free surface and thereby transfer
its whole energy to microbubbles via multiple excitations. This full energy
transformation offers a new way to estimate integral \emph{shockwave energies}.
The idea consists in integrating the energies of all microbubbles, which can be
inferred from statistical measures of their densities and maximal volumes
$V_{max}$ via $E=V_{max}(p_\infty-p_V)$ \cite{Rayleigh 17}, where $p_\infty$ is
the ambient static pressure and $p_V$ is the vapor pressure. Applying this
method to five concrete samples showed that 50-70\% of the initial electrical
energy (200 mJ) is contained in the mist of microbubbles associated with the
spark shock. It follows that 50-70\% of the spark energy is radiated in a
shockwave, and we expect to find the remaining 30-50\% in the main bubble; a
prediction which is consistent with the measured maximal main bubble volume
corresponding to a bubble energy of 70 mJ. The actual precision was limited by
the optical resolution and non-vanishing contact surfaces (injector tube,
electrodes), and could be improved in forthcoming studies.

\emph{Spherical Collapse} $-$ We shall now investigate the collapse of bubbles
\emph{centered} in drops ($\varepsilon=0$), since their spherical symmetry
allows a direct comparison to bubbles in infinite volumes, governed by
Rayleigh-Plesset's theory \cite{Rayleigh 17,Plesset 49}. Bubbles in drops were
found to collapse faster in excellent agreement with a theoretical extension
derived in the following. Note that gravity does not play a direct role in
Rayleigh-Plesset's theory, nor in our extension, but its absence allowed the
realization of isolated drops with centimetric diameters.

In the present approach we neglect surface tension, liquid compressibility,
viscosity and mass transfer across the bubble boundary. We further assume a
uniform and constant temperature, and take the pressure inside the bubble as
the vapor pressure $p_V$. The ensuing simplified Rayleigh(-Plesset) equation
for a spherical bubble with radius $R_b(t)$ in an infinite liquid reads
\begin{equation}\label{rayleigh-plesset}
  \frac{-\Delta p}{\rho}=\frac{3}{2}\dot{R_b}^2+R_b \ddot{R_b}
\end{equation}
where $\rho$ is the liquid density and $\Delta p\equiv p_\infty-p_V$ with
$p_\infty$ being the liquid pressure taken far from the bubble. Integrating
(\ref{rayleigh-plesset}) provided the initial conditions $R_b(0)=R_{b,max}$ and
$\dot{R_b}(0)=0$ yields the radius $R_b(t)$ of the collapsing bubble. The
so-called \emph{Rayleigh collapse time} $T_{Rayl}$, such that
$R_b(T_{Rayl})=0$, is given by $T_{Rayl}\approx 0.915
R_{b,max}\sqrt{\rho/\Delta p}$\, \cite{Rayleigh 17}. To compare different
bubble collapses, we present their functions $R_b(t)$ in normalized scales
where $T_{Rayl}$ and $R_{b,max}$ equal one. In these scales, $R_b(t)$ can be
shown to be independent of $\Delta p$, $\rho$ and $R_{b,max}$. To assert the
validity of the above model for the present situation, we carried out extensive
ground-studies with millimetric bubbles in extended water volumes. The data
revealed excellent agreement with the theoretical prediction (Fig.\ \ref{fig
4}, dashed line and squares), thus justifying the above assumptions.

\begin{figure}[h]
  \includegraphics[width=8.3cm]{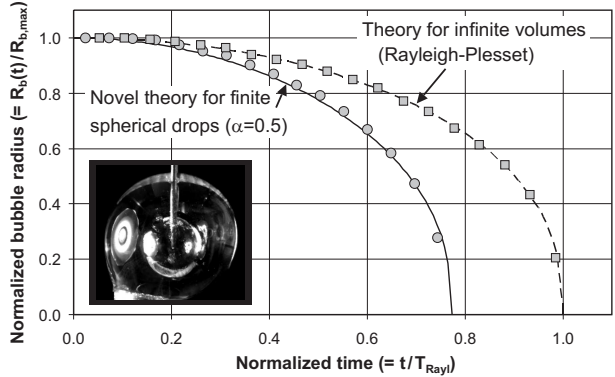}
  \caption{Bubble collapse.
  (dashed line) Rayleigh-Plesset theory (\ref{rayleigh-plesset}).
  (squares) Ground-experiment in extended water volumes.
  (solid line) Modified theory for spherical
  drops, $\alpha$=0.5 (\ref{modified rayleigh-plesset}).
  (circles) Microgravity-experiment in drops.
  Errors are given by the size of the data points.}
  \label{fig 4}
\end{figure}

In microgravity, we studied bubbles centered in drops with a relative radius
$\alpha=0.50$ ($R_{b,max}$ = 4.07 mm, $R_{d,min}$ = 8.15 mm, cabin pressure
$p_\infty=80\,000\ \text{Pa}$, spark energy = 200 mJ). The bubble collapses
were sampled at 50\,000 frames/s with an optical resolution of 70 $\mu$m. Given
this high spatiotemporal resolution, the results from 5 experimental runs were
indistinguishable and ensured reproducibility. The collapse time was measured
as $330\pm 10 \mu$s and the evolving bubble radius $R_b(t)$ was reconstituted
from the images using a model of optical refraction by a water sphere
\cite{FS}. Plotted in normalized scales, the microgravity data reveals a
remarkable \emph{shortening of the collapse time} (Fig.\ \ref{fig 4}, circles).

In order to explain this shortening, we shall now derive an equation of motion
for cavitation bubbles of radius $R_b(t)$ centered in spherical drops of radius
$R_d(t)$, using the above assumptions. Incompressibility links $R_b(t)$ and
$R_d(t)$ and conditions the radial velocity $\dot{r}$, as seen by integrating
$div\ \vec{v}=0$ along a radial axis,
\begin{eqnarray}
  R_d(t) & = & \left(R_{d,min}^3+R_b^3(t)\right)^{\frac{1}{3}}
  \label{radius relation} \\
  \dot{r} & = & \dot{R}_b R_b^2/r^2
  \label{velocity field}
\end{eqnarray}
For the sake of physical interpretation, we derive the equation of motion from
energy conservation. The kinetic energy results from integrating the energy
density $\frac{1}{2}\rho\dot{r}^2$ over the whole liquid volume and the
potential energy is given by integrating the pressure work $p dV$ over the
volume traversed by the bubble surface \emph{and} the drop surface. Using
(\ref{radius relation}) and (\ref{velocity field}),
\begin{equation}
  E_{kin} = 2\pi\rho \dot{R}_b^2 R_b^3(1-\lambda),\ \
  E_{pot} = \frac{4\pi R_b^3}{3}\Delta p\quad
\end{equation}
with $\lambda(t)\equiv R_b(t)/R_d(t)$. Invoking the conservation law
$\dot{E}_{kin}$+$\dot{E}_{pot}=0$ and dividing by $4\pi\rho\dot{R}_b R_b^2$
yields the equation of motion for bubbles centered in drops,
\begin{equation}\label{modified rayleigh-plesset}
  \frac{-\Delta p}{\rho} = \frac{3}{2}\dot{R}_b^2+R_b\ddot{R}_b
  -2\lambda\dot{R}_b^2-\lambda R_b\ddot{R}_b+\frac{1}{2}\dot{R}_b^2\lambda^4
\end{equation}
where the term in $\lambda^4$ was obtained by substituting $\dot{R}_d$ using
relation (\ref{radius relation}). This equation comprises the simplified
Rayleigh-Plesset equation (\ref{rayleigh-plesset}) and three novel corrective
terms, which vanish as $R_d\rightarrow\infty$. Up to the factor
$4\pi\rho\dot{R}_b R_b^2$ the two terms in $\lambda$ subtract the change of
kinetic energy associated with the volume outside the drop ($R_d(t)<r<\infty$),
which needs to be removed from the infinite medium of Rayleigh's model. The
remaining corrective term in $\lambda^4$ accounts for the change in kinetic
energy due to the variation of the drop's radius. This term becomes negligible
for relative bubble radii $\alpha\ll 1$ since $\lambda(t)<\alpha\ \forall\ t$.
Integrating (\ref{modified rayleigh-plesset}) with the initial conditions
$R_b(0)=R_{b,max}$ and $\dot{R_b}(0)=0$ gives the collapsing bubble radius
$R_b(t)$, which is in excellent agreement with the microgravity data
(Fig.~\ref{fig 4}, solid line). This proves that the shortened lifetime is
entirely due to the finite spherical drop volume, and validates the derived
equation of motion for bubbles centered in spherical drops (\ref{modified
rayleigh-plesset}).

\begin{figure}[h]
  \includegraphics[width=8.3cm]{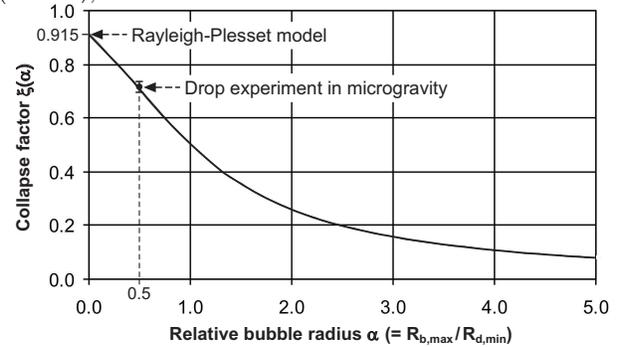}
  \caption{Collapse factor $\xi(\alpha)$ as a function of the relative bubble
  radius $\alpha$. $\xi(\alpha)$ is directly proportional to the collapse time
  $T_{collapse}$ if $\rho$, $\Delta p$ and $R_{b,max}$ are fixed.}
  \label{fig 5}
\end{figure}

Finally, we apply the validated model to predict the collapse times of bubbles
with arbitrary relative radii $\alpha$. The energy conservation
$E_{kin}+E_{pot}=E_0$ can be solved for $\dot{R}_b$, if $E_0$ is identified
with the potential energy at maximal bubble radius $R_{b,max}$,
\begin{equation}\label{collapse speed}
  \dot{R}_b = -\sqrt{\frac{1}{1-\lambda}\frac{2\Delta p}{3\rho}
  \Big(\frac{R^3_{b,max}}{R_b^3}-1\Big)}
\end{equation}
Integrating this equation from $R_b=R_{b,max}$ to $R_b=0$ yields an analytical
expression for the collapse time,
\begin{eqnarray}\label{collpase time}
  && T_{collapse} = \xi(\alpha)\,R_{b,max}\sqrt{\frac{\rho}{\Delta p}} \\
  && \xi(\alpha) \equiv \sqrt{\frac{3}{2}}\,\int_0^1
  \Big(1-\frac{s}{(\alpha^{-3}+s^3)^{1/3}}\Big)^{\frac{1}{2}}
  \Big(\frac{1}{s^3}-1\Big)^{-\frac{1}{2}} ds \quad\nonumber
\end{eqnarray}
where $s$ substitutes $R_b/R_{b,max}$. The collapse time depends on $\alpha$
through a \emph{collapse factor} $\xi(\alpha)$ plotted in Fig.\ \ref{fig 5}.
For infinite liquids ($\alpha=0$) we consistently recover the \emph{Rayleigh
collapse factor} $\xi(0)=0.915$ \cite{Rayleigh 17}. $\xi(\alpha)$ decreases
monotonically with increasing relative bubble radius $\alpha$, and tends to 0
for $\alpha\rightarrow\infty$. Even though this limit could not be probed in
the microgravity time available, it seems non-physical for it violates the
model assumptions. In particular, for $\alpha\gtrsim6$ the drop's surface
velocity $\dot{R}_d(t)$ would approach the speed of sound for a significant
fraction of the collapse phase [use (\ref{collapse speed}), (\ref{radius
relation})], thus implying a strong pressure drop and preventing further
acceleration.

\emph{Conclusion} $-$ Microgravity conditions proved essential in the study of
cavitation bubbles inside centrimetric water drops. In particular, (1) we
obtained the first simultaneous visualizations of both cavity-induced liquid
jets, and discovered a remarkable splash broadening and missing crown formation
on spherical surfaces. (2) Reflected shockwaves led to a strong form of
secondary cavitation and offer an original way for the measurement of shockwave
energies inside isolated volumes. (3) Our Rayleigh-Plesset-like model for
bubbles in drops is in excellent agreement with specific experiments and
allowed a general prediction of bubble collapse times in drops. Forthcoming
microgravity studies could fruitfully address the limit of large bubble radii
($\alpha\gg 1$), or focus on direct shockwave visualizations.

We thank the ESA for having pursued this research and the \emph{Swiss NSF} for
Grant No.~2000-068320.

\end{document}